# Safe and Secure Smart Home

Shubham Tripathi, Vanaparthy Akshith, Shivansh Walia, Tejas Iyer, Shubham

## Abstract

This project presents an implementation and designing of safe, secure and smart home with enhanced levels of security features which uses IoT-based technology. We got our motivation for this project after learning about movement of west towards smart homes and designs. This galvanized us to engage in this work as we wanted for homeowners to have a greater control over their in-house environment while also promising more safety and security features for the denizen. This contrivance of smart-home archetype has been intended to assimilate many kinds of sensors, boards along with advanced IoT devices and programming languages all of which in conjunction validate control and monitoring prowess over discrete electronic items present in home.

We contrived this system with the help of contemporary release of Cisco Packet tracer which was used for analytical and simulation purposes, this also allowed us to do on-spot design, analysis, testing and cognition of the systemized network. The above mentioned system has been implemented with a ton of enriched security visage which include Radio Frequency Identification (RFID)-based door access control system, burglary detection system operating through motion detection with webcams, fire recognition and control system compromising fire monitoring, smoke detection, fire sprinklers along with window control and siren, this arrangement also comprises of water level monitor for enclosures in conjunction of water sprinklers to assure safety of residents. Hereby mentioned system has also been intended to be operated and monitored subordinately with the use of mobile devices providing the residents with elevated control and added on convenience  for their household.

This research paper concludes with the breakdown of the results of this study and converse regarding their  connotations for further analysis and exploration and research and development of smart home technology. The strengths of above-mentioned system along with some of its limitations have been featured also recommendations and areas to improve upon have been bought to light. This research paper attempts to bring light to the modern need for amalgamating enriched and enhanced security and safety visage into smart home builds. Our planned and presented arrangement is set to offer a comprehensive system for  homeowners and residents looking to manage their home environment remotely while also establishing security and safety for house and its denizen. This research tries to offer valuable information for the architecture and execution steps for implementing smart and secure home systems with enhanced safety and security features.

## Introduction

The theory about the implementation of smart home and automating everyday tasks with a system where daily devices are linked and are controlled through the internet from any place has been very prominent since last two decades and with the improvement of technology and new innovations in conjunction to the increasing availability of IoT devices many homeowners are in a search for better and better ways to control their home environment remotely, While these smart homes may offer many benefits such as convenience, resource efficiency and easy monitoring, they also bring forward many safety and security challenges which are required to be addressed, therefore the focus of this research paper has been to architect a smart home design with enhanced security functionalities using IoT-based inventions which should provide homeowners and residents with a greater authority over their home's environment while also establishing safety and security for home and its denizen.

The most common problems which are being faced in designing such a smart home system involves security risks, complications, and interoperability, these security risks arise from the collaboration and interlinkage of multiple devices within the network, as it can make the system more susceptible to attacks like hacking and malware. Complexity is another roadblock as the count of devices used to make the system and the bulk of data generated by them can be astounding and incredible to handle making the administration and maintenance of very overwhelming. Interoperability actually refers to the ability of connected devices to be able to work seamlessly, as the devices used in the system can have many different properties and can be from many differences manufacturers while also containing different protocols, establishing compatibility and interoperability could be very challenging. Through our reviews we found out that the most of the smart home security systems include firewalls, encryption techniques and user authentication, while these all do offer a level of security, but still they don't look into all the security issues associated with smart homes, as Firewalls and encryption can still be bypassed and user authentication can be compromised through phishing attacks or by cracking weak passwords.

The handicap of above-mentioned approaches is that they generally fail to provide an all-inclusive solution addressing all the security challenges which accompany smart home systems. As they mostly focus on securing individual devices instead of addressing security issues of system as a whole, also many of these systems actually lacks the ability to be controlled remotely, denying most of the accessibility and control that homeowners desire. This proposed system is architect with updated and enhanced security and safety features to protect the home environment for making sure the safety and security of its denizen. The advanced features of this system include Radio Frequency Identification (RFID)-based door access control system and also a similar garage system, burglary detection system , water level monitoring, fire control system, smoke detection, temperature monitoring and moderation system and webcams, these enriched features are not only upgrading the automated safety and security mechanisms but are also going to provide early admonition for potential threats like burglary, fire and water damage.

The uniqueness in this research paper is all-inclusive and extensive approach to architecture a smart home system with enriched security, this proposed system is for addressing the security problems associated with smart homes by integrating a wide range of features which work together in a smooth and uninterrupted manner. This system is architect to be managed and governed remotely, also providing homeowners with elevated authority and convenience for their home environment, on top of this the simulation platform used for this research i.e. Cisco Packet Tracer (CPT) allows for rectification of the system's security features in instants, which assure and establish that it functions well for security, safety and convenience. In a nutshell, this research paper features the importance of having a conjunction of enriched safety and security features into smart homes. The System mentioned is to offer an all-inclusive and complete solution for householders finding solutions to have authority over their home environment remotely while also assuring safety and security for the home and its denizen. This research serves worthwhile insights into the architect and implementation of this smart home system with enriched security features, this research also suggests areas of future analysis, experimentation and innovation.

## Literature Review/Related Work

In the last two decades use of IoT devices for implementing smart homes have been on a rise. The concept of smart home moves around the usage of technology for automating daily home tasks and for administrating and governing multiple appliances, devices and systems for home, as the new innovations in field of communication technology are rising and availability of IoT devices is increasing the implementation of smart home design is getting easier. Several methods for implementing smart home with enhanced security have already been proposed. A popular one among these approaches is to use sensors and cameras to monitor home environment, for instance Hossain at el. (2018) suggested a system which uses sensors and cameras to detect motion and monitors home environment which in case of security breach alerts homeowners. After this another popular solution is to use Radio Frequency Identification (RFID) based access control systems this involves using RFID tags to limit access of areas of home, this system can be used to grant an access to only authorized people Wang et al. (2017), proposed one such system. These approaches do have a drawback though and it's that they don't provide an all-inclusive solution for the problem of security, as use of cameras and sensors do help in detecting intruders but they don't address safety hazards and fire and water damage, similarly the RFID based system don't address possibility of cyber-attacks.

Lin et al. (2019) proposed a similar system to that of ours in his study he was incorporating multiple IoT devices for monitoring and administrating home appliances but the system still was in deficit of modern security vices like fire and smoke detection and control systems. Another similar research by Chen et al. (2018) proposed smart home system which was using machine learning algorithms which were to detect abnormal activities in home environment, while such an approach successfully detects security breaches but it wasn't a comprehensive solution, the existing literature do talks about many approaches to architect smart homes with enriched security, though these approaches were useful but they all were having some limitations which were required to be addressed. Our system proposes an all-inclusive safe and secure smart home which had incorporated advanced security features to deliver a solution. The proposed system can also be remotely operated and monitored, which again enables enhanced convenience and control over the home environment.

The literature review for this study had articulated a significant divide in already conducted research relating to architecture and application in operation of smart home systems along with enriched safety and security visage. This work dwells upon this through proposed extensive and comprehensive elucidation which again presents the residents with tonnes of features which are meant to deliver enriched and enhanced security and safety. The analytics represents that this arrangement hereby discussed has and is headed to perform great in terms of providing residents with security, safety and convenience. Through these analytics we were also able to confirm that the enhanced and enriched features like controlled access of doors and Garage with auto closing and monitored windows in conjunction with motion detection system with webcam around the home established security for denizen, And automated features like water control for enclosures, fire control system, and temperature monitoring system helped in fabricating a safe home providing initial warning and safeguard from threats like theft, fire and water damage. Furthermore the demonstrated system was also successfully regulated and governed remotely, which again ensures more authority and freewill over the home and home environment for the homeowners.

# Study Area and Data Resources

The study area for our research is architect and execution of a smart home system which provides enriched security features, this research brings light on delivering a prototype system with IoT devices and sensors for governing electronic devices in home.

Data resources used in this study involves usage of Cisco Packet Tracer, which has provided a platform for testing and simulation of the network in real time, this system incorporated various sensors boards programming languages and IOE devices which are used for governing electronic devices in home, moreover dataset used for this study incorporates sensor readings like temperature , moisture levels, motion detection data and water level readings, which are being collected by the sensors and are stored in a database for further analysis, the system also uses generated logs for monitoring performance and detecting potential threats and also stores alerts in case of security breaches.

# Methodology

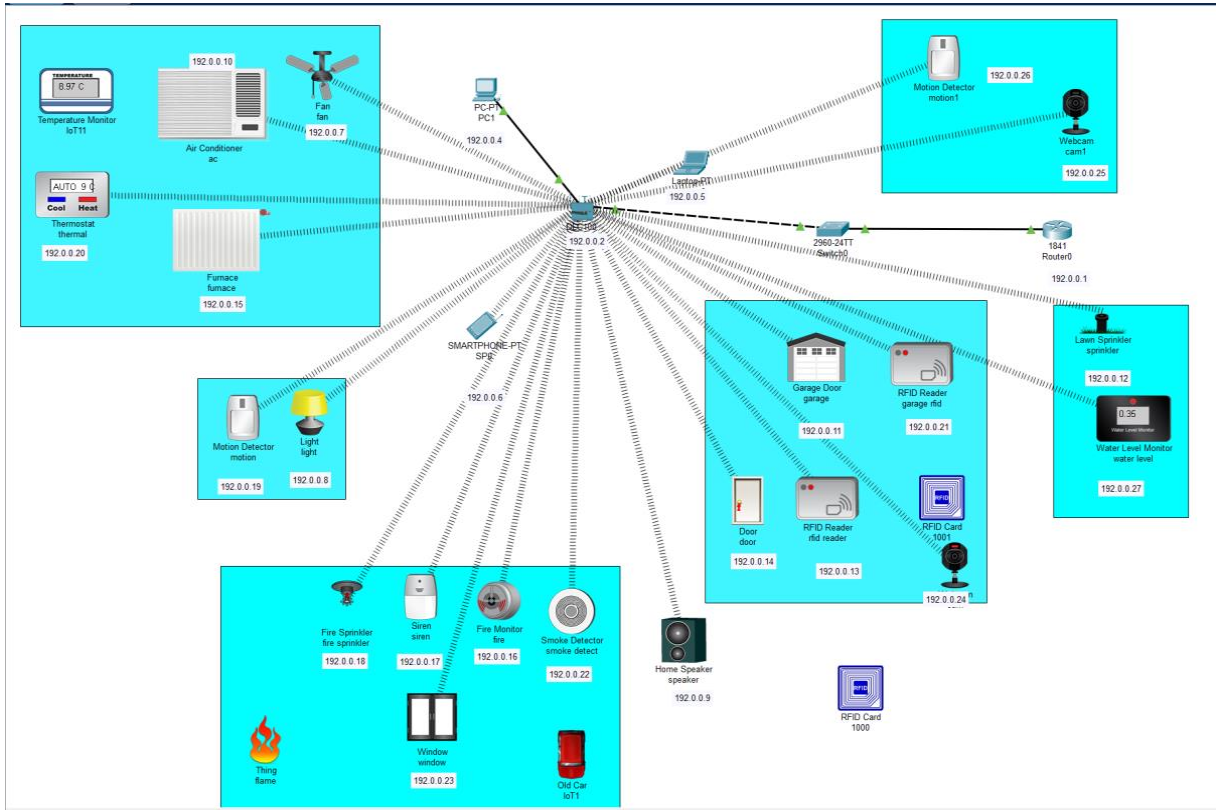

Figure 1

Figure 2

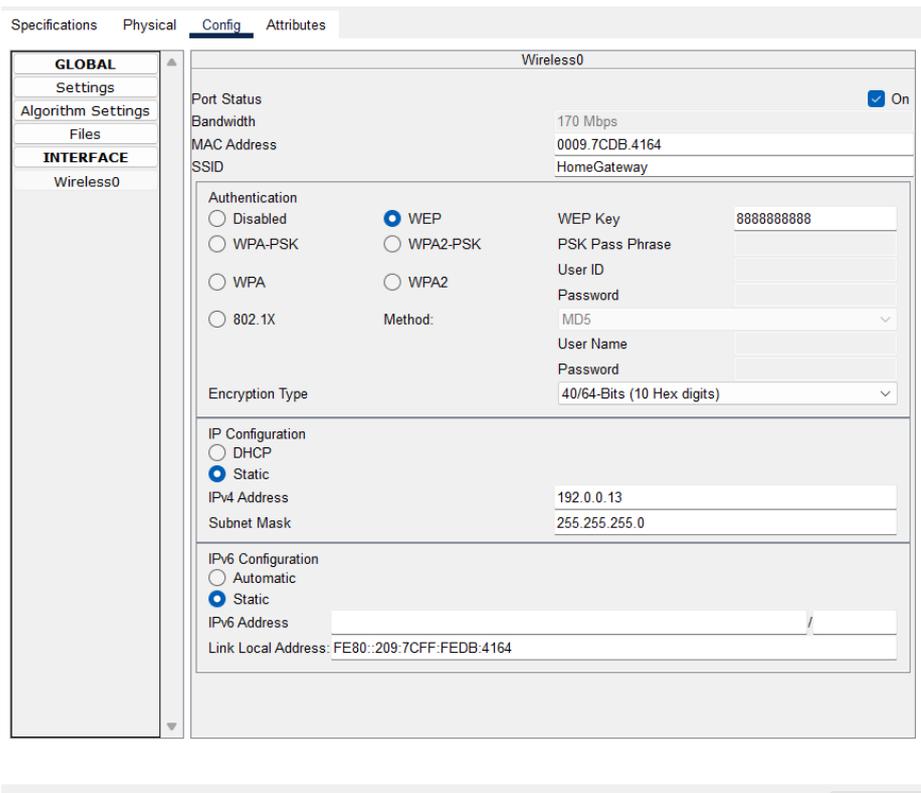

Figure 3

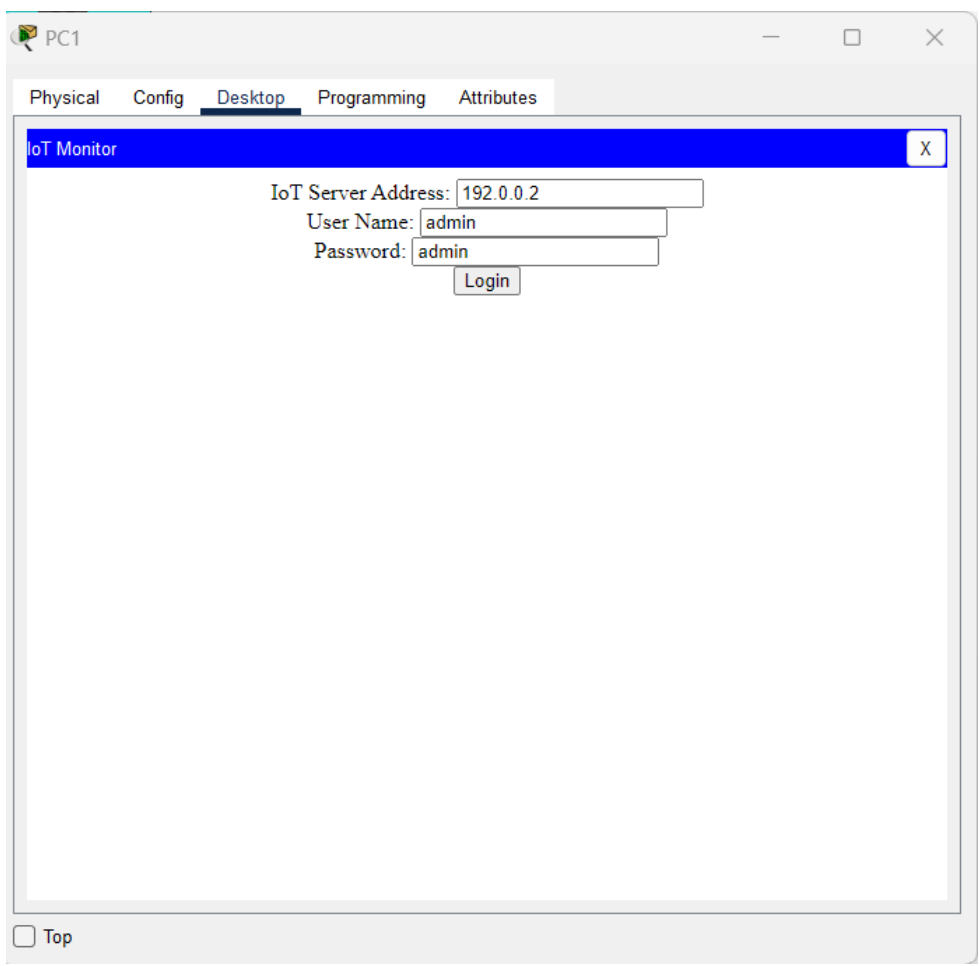

Figure 4

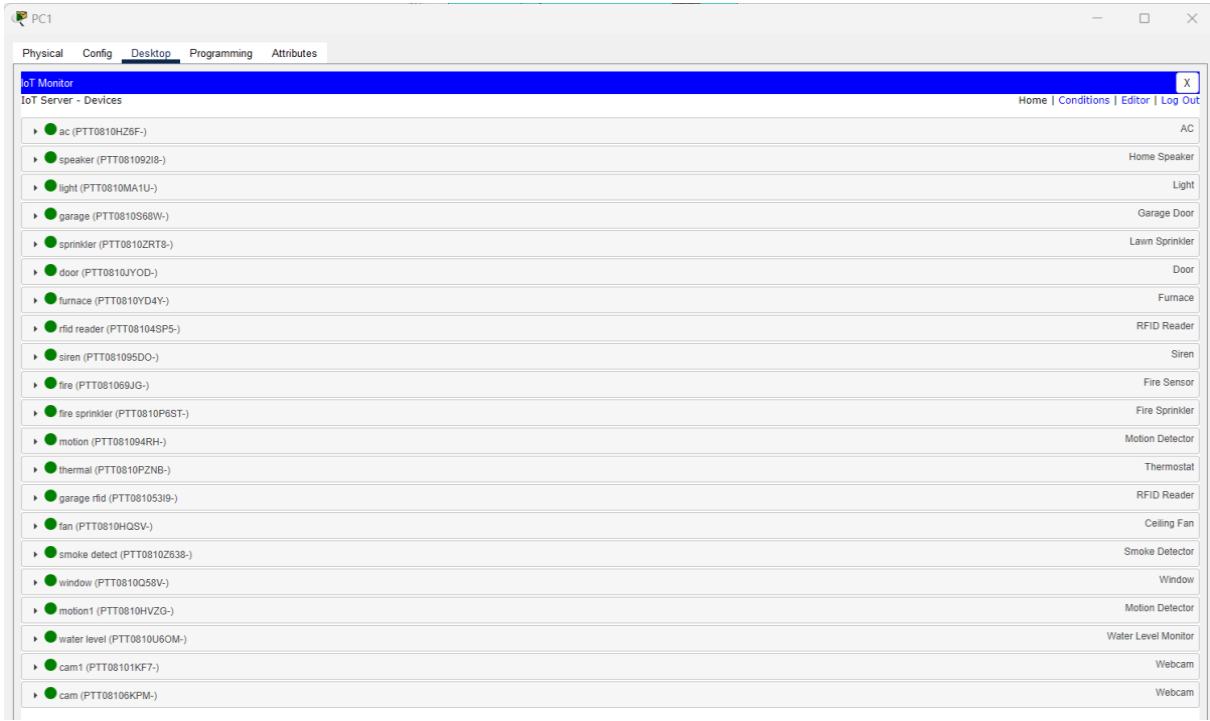

Figure 5

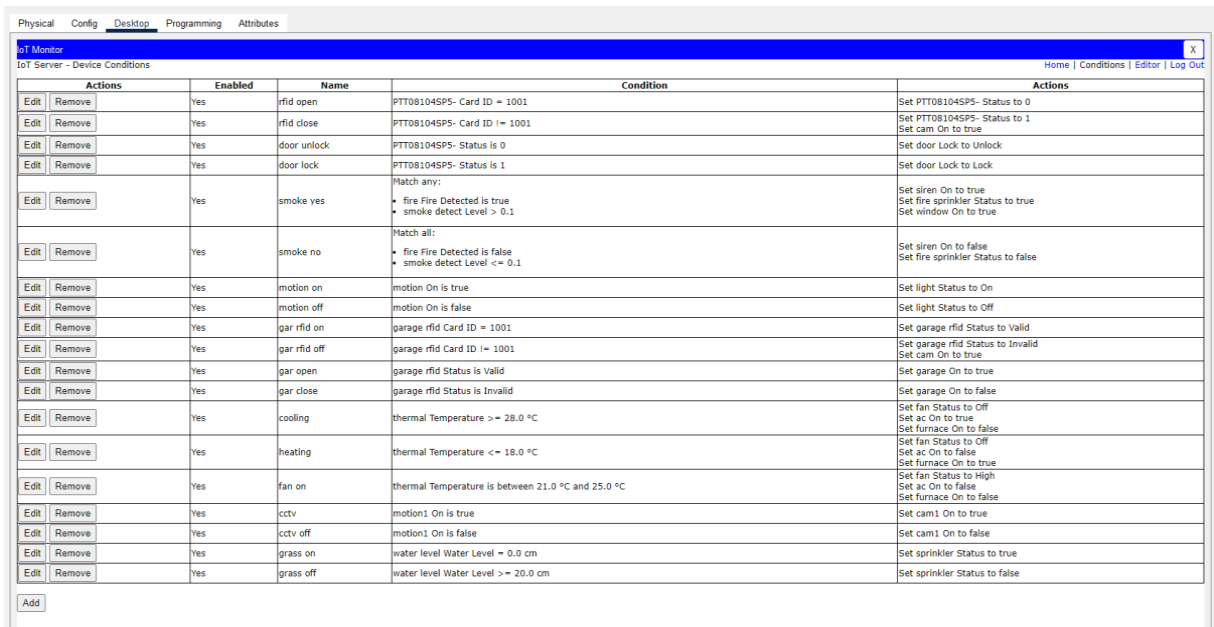

Figure 6

To implement smart and secure home utilizing Cisco packet tracer 8.2 version. The figure 1 represents the smart and secure home design that communicate with each other utilizing wireless medium. We have assigned the IP address to the router and IOT devices. Then we set authentication WEP to our Home Gateway. While connecting the IOT devices we set the default gateway to our router IP address and IOT server as Home Gateway shown in figure 2.Enter your Home Gateway authentication password to our IOT devices shown in figure 3. All the devices are now registered on our Home Gateway. These devices are now accessible through a pc connected to the Home Gateway. Login using home Gateway address to configure IOT devices shown in figure 4.

These IOT devices can be accessed now shown in figure 5.We have RFID reader which validates the RFID card number to open main door and garage door. There is fire safety system where when fire monitor or smoke detector detects fire or smoke respectively and sets off fire sprinklers, sirens and opens the window . The water level monitor detects the water in grass and turns on the lawn sprinkler accordingly. There are motion detectors setup ,when motion detected lights and cameras turn on. For temperature control we have used a thermostat which turns on AC when room temperature is greater than 28 degrees Celsius and turns on Furnace when room temperature is less than 18 degrees Celsius. The conditions for how these devices communicate with each other can written using python or JavaScript or use conditions menu in IOT Devices shown in figure 6

## Results and discussion

This researched system is an implementation of safe and secure smart home has enriched the security by adding Radio Frequency Identification (RFID)-based door and garage access control systems and burglary detection system operating through motion detection with webcams and creates a safe environment through water level monitoring, fire control system, smoke detection, temperature monitoring and moderation system.

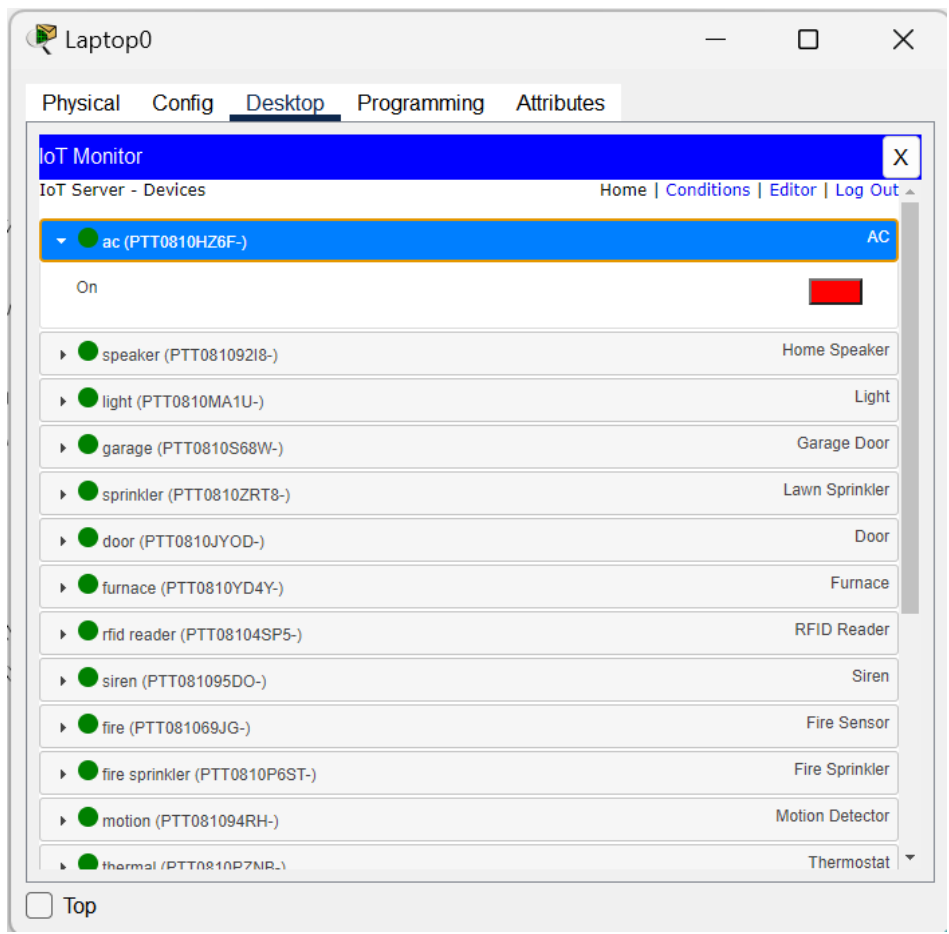

Figure 7: Shows the working of IoT devices.

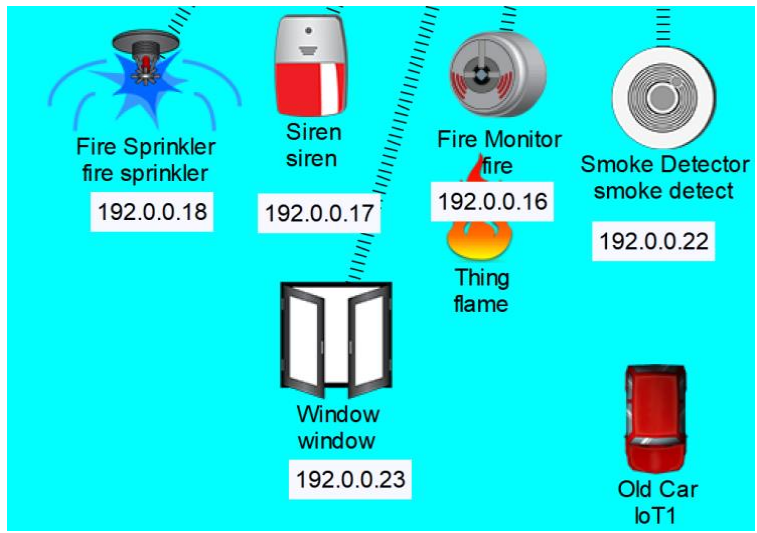

Figure 8: Shows activation of siren, sprinkler and opening of window as fire monitor is triggered.

# Performance evaluation/validation

The performance assessment for this smart home system with enriched security vices has been carried out with the help of various assessment and measurement procedures to assure that the system has been contesting the desired exigences, point of assessment that were used in this research includes response times, system start time, and network latency, the response times had been measured as the time duration for which the system was operational without any downtime which was found to be above ninety-nine percent across the evaluation time, the network latency had been measured as the time consumed by the system to transfer data from one device to another device and we found that it was less than ten milliseconds. To verify the performance of the solution these three tests were performed:

1. System was evaluated on its capability to detect and counter to security breaches, here the system was exposed to multiple security attacks and it was able to respond to them effectively.
2. System's capability to administer and monitor various electronic devices in home was evaluated upon various factors like maintaining temperatures, lighting conditions, and security cameras, and was eventually found to be able to control them effectively.
3. System's capability to be able to control and monitor the home environment remotely was evaluated using smartphone and it was successfully verified.

# Conclusion

Through this research paper we exhibited an architect for smart, safe and secure home system based on IoT-based innovations with enriched security features. The system which is being reflected allows management and governing of multiple different electronic devices present in home through remote connectivity using end devices. This system has been implemented with the help of Cisco Packet Tracer(CPT) which is a platform allowing for simulation and testing of network in an instant.

For this research project we firstly discovered all the major problems faced by conventional and existing home security systems and then listed all the limitations of those solutions. We had also examined the existing research papers on the topic in hand and also tried to list down their shortcomings of their approach. Our proposed system overcomes these limitations by including advanced features such as RFID based door access control, motion controls, webcams, water level monitoring, temperature monitoring, motion detection using webcams, smoke detection, fire detection and control with alerts and an RFID-based garage system.

For our work we also organized performance assessments and validation analysis on the proposed system to confirm its performance and effectiveness, from these the outcomes of assessments we got reflects that our system is capable of fulfilling enriched security and safety measures for home environment.

Subsequent work on this could be focused upon advancing the system to be able to incorporate extra progressive features like voice detection and assistant along with smart automations to enrich the system's comprehensive performance and convenience for the users. Current barriers of our final solution include Requiring an stable internet connection and Probable risk of data breaches, to overcome these limitations, further studies could be focused on evolution of progressive security measures to conserve the system from potential cyber threats. In the end I would like to state that our proposed smart home system which includes enriched security visage delivers a progressive solution for the problems which inhabits pre-existing home security systems, this system also delivers remote administration and governing capabilities over home network along with enriched security features, a beforehand warning and safety against possible dangers, enhancing the safety and convenience for the home environment.  From our study we have demonstrated potential of technology based on IoT for architect of smart homes with advance security features.

# Acknowledgement

Firstly, we would like to appreciate and express our gratitude to everyone who have in any way helped us in this project, Including Dr. Yogesh pal and Dr. Ashish kumar for their supervision, advices, guidance, support and their treasured insights throughout the development cycle of this project. We are also thankful to our colleagues who helped us with ideas, recommendations and assessment during the evolution and analysis phases of our system. We are also grateful and would like to appreciate the development team of Cisco Packet Tracer for the software on which we are able to simulate and represent our project and Network skills. In absence of these tools the advancement, analysis and testing of the proposed project wouldn't had been possible. And definitely in the end we would like to appreciate our friends and family for their intense and regular support and encouragement throughout the development process. Support from them had been vital and has always motivated us to be preoccupied with the development of this project.